\documentclass[a4paper,twocolumn]{esapub}
\usepackage{ifluatex,ifxetex} 
\ifluatex
\usepackage{fontspec}
\else\ifxetex
\usepackage{fontspec}
\else
\usepackage{times}
\usepackage{amsmath}
\usepackage{hyperref}

\fi

\usepackage[numbers]{natbib}
\usepackage{graphicx}

\title{A systematic ranging technique for follow-ups of NEOs detected with the Flyeye telescope}
\author{Michael Fr\"uhauf}
\affil{ESA/ESOC, Robert-Bosch-Str. 5, 64293 Darmstadt (Germany), Email: Michael.Fruehauf@esa.int}
\author{Marco Micheli}
\affil{ESA/NEO Coordination Centre, Largo Galileo Galilei 1, 00044 Frascati (Italy), Email: Marco.Micheli@esa.int}
\author{Toni Santana-Ros}
\affil{ESA/ESOC, Robert-Bosch-Str. 5, 64293 Darmstadt (Germany), Email: Toni.Santana-Ros@esa.int}
\author{R\"udiger Jehn}
\affil{ESA/ESOC, Robert-Bosch-Str. 5, 64293 Darmstadt (Germany), Email: Ruediger.Jehn@esa.int}
\author{Detlef Koschny}
\affil{ESA/ESTEC, Keplerlaan 1, 2201 AZ Noordwijk (The Netherlands), Email: Detlef.Koschny@esa.int\newline
Lehrstuhl f\"ur Raumfahrttechnik, TU Munich, Boltzmannstr 15, 85748 Garching (Germany)}
\author{Olga Ram\'irez Torralba}
\affil{ESA/ESOC, Robert-Bosch-Str. 5, 64293 Darmstadt (Germany), Email: Olga.Ramirez.Torralba@esa.int\newline
TU Delft, Mekelweg 5, 2628 CC Delft (The Netherlands)}

\begin{document}


\maketitle

\begin{abstract}
When new objects are detected in the sky, an orbit determination needs to be performed immediately to find out their origin, to determine the probability of an Earth impact and possibly also to estimate the impact region on Earth. ESA's Flyeye telescope is expected to revolutionize the effort of predicting potential asteroid or deep space debris impact hazards due to the expected increase of near-Earth object discoveries. As the observed orbit arc for such an object is very short, classical Gaussian orbit determination cannot be used. We adopt the systematic ranging technique to overcome the lack of information and predict a region of the sky where the body can most likely be found. We also provide a detection probability for follow-up observations and investigate potential follow-up telescopes for the Flyeye telescope.
\end{abstract}

\section{Introduction}

In recent years, many near-Earth objects (NEOs) have been detected, leading to a current (2019, Jan.) number of more than 19\,300 known objects, listed on ESA's NEO Coordination Centre website\footnote{http://neo.ssa.esa.int\label{FN:neossaesaint}}. Due to an increasing number of survey telescopes, the discovery rate has risen over the past decades, reaching a value of 2033 objects in 2017 and 1822 objects in 2018. The monthly NEOCC newsletter\textsuperscript{\ref{FN:neossaesaint}} provides regular updates of the newly detected objects. The lower rate of 2018 can be explained by bad weather conditions in Hawaii and an unplanned downtime of Pan-STARRS \cite{ramanjooloo2018the}, showing that the discovery of NEOs shall not rely on a single observatory, but on several independent telescopes, spread all over the globe.

ESA will contribute to the task of searching for uncatalogued NEOs with the Flyeye telescope, a telescope splitting the large field of view ($6.7$\;deg by $6.7$\;deg) by use of a prismatic faceted mirror onto 16 individual cameras \cite{cibin2019the}. The telescope will allow the coverage of almost a full hemisphere every two  observing nights. Each area of the sky is visited four times, collecting information on the possible movement of objects brighter than $21.5$\,mag. When a new object is found, immediate orbit determination needs to be done to filter potential impactors from passing NEOs and main-belt asteroids, estimating the impact probability and a potential impact region on Earth. Examples of impactors are the asteroids 2008\;TC3, 2014\;AA and 2018\;LA, which all were discovered by R. A. Kowalski as part of the Mt. Lemmon Survey (G96) and impacted less than a day after they were detected \cite{MPC2008, MPC2014, MPC2018}.

\begin{figure*}
    \centering
    \includegraphics[width=1.00\textwidth]{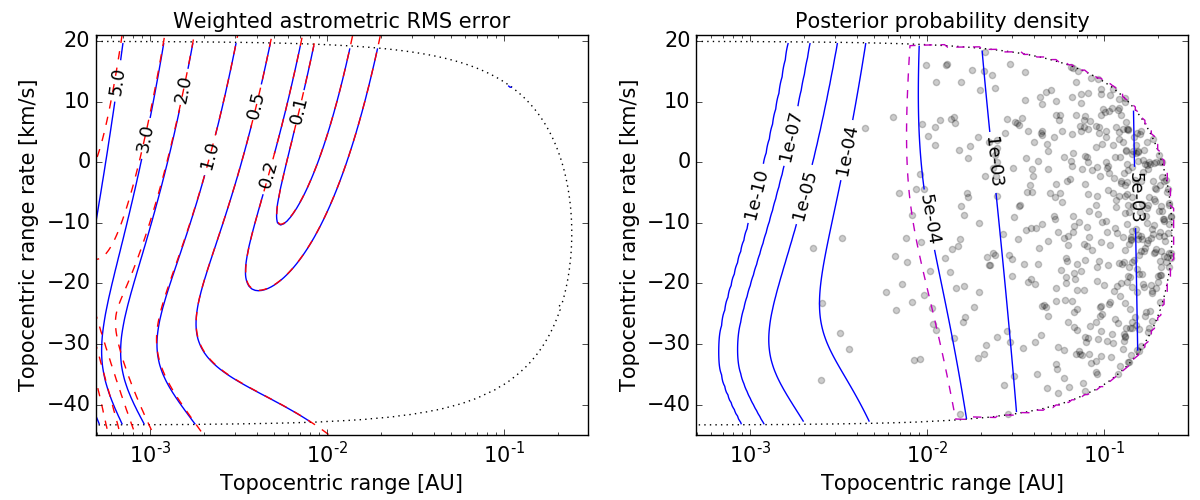}
    \caption{The left plot shows the weighted astrometric root mean square (RMS) error of the object with the temporary designation ZW04D95 (blue solid contour lines). The overlaying contour map (red dashed contour lines) is the result of D. Farnocchia (personal communication, 2018). The right plot shows the corresponding posterior probability density (blue solid contour lines), the 95\% confidence region (magenta dashed line) and 500 Monte Carlo generated samples (black dots). The black dotted line is the boundary between elliptic and hyperbolic orbits.}
	\label{fig:systematicranging_ZW04D95}
\end{figure*}

For a newly discovered object, the observed orbit arc is usually very short \cite{milani2004celestial}, with just four reliable observables: right ascension, declination, and the rate of change of right ascension and declination. For this reason, classical Gaussian orbit determination cannot be applied until further follow-up observations are obtained. However, without a good knowledge of the orbit, follow-up observations might point to wrong celestial coordinates and miss the object, leading to a vicious circle. In order to find the object a few or even many hours later, it is necessary to put some constraints on the region of the sky to observe. For this goal, we are implementing the technique of systematic ranging, described in Chapter\;\ref{Sec:SystematicRanging}, in our system, where a raster scan in the space of topocentric range and range rate is performed. By assigning a probability density to this space, we generate in Chapter\;\ref{Sec:bestfollowuppointing} a sample cloud on the sky, which can be used by observers to select a telescope pointing. In Chapter\;\ref{Sec:PointingTelescope} we calculate a promising pointing for a specific telescope that increases the detection probability. Furthermore we investigate the detection probabilities for a number of collaborating follow-up stations of the Flyeye telescope.

\section{Systematic ranging} \label{Sec:SystematicRanging}

For a newly discovered object, we have a set of $n$ observations, the so-called tracklet, consisting of right ascensions $\alpha$, declinations $\delta$ and potentially apparent magnitudes $m$. With this data, rates of right ascension $\dot{\alpha}$ and rates of declination $\dot{\delta}$ can be determined, leading to the so-called attributable $\mathbf{A}=(\alpha,\delta,\dot{\alpha},\dot{\delta})$ \cite{milani2004celestial}. However, an orbit is fully defined by 6 parameters, which means two additional parameters have to be guessed.

We follow the systematic ranging technique as described in \cite{chesley2004very} and \cite{farnocchia2015systematic}, using the topocentric range $\varrho$, the distance between the observer station and the observed object, and the range rate $\dot{\varrho}$ as intuitive 5th and 6th parameter. The lack of information is overcome by systematically scanning a ($\varrho$,$\dot{\varrho}$)-grid, in contrast to the Monte Carlo based statistical ranging \cite{virtanen2001statistical}, where $\varrho$ and $\dot{\varrho}$ are randomly sampled. For each grid point with fixed $\varrho_i$ and $\dot{\varrho}_j$, a best-fit attributable $\mathbf{A}_{ij}(\varrho_i, \dot{\varrho}_j)$ for the first observation epoch and a corresponding orbit can be determined by minimizing the cost function \cite{jehn1989orbdet}:
\begin{equation}\label{Eq:CostFunction}
    \boldsymbol\nu^T W \boldsymbol\nu = \textrm{minimum}
\end{equation}
where $\boldsymbol\nu=(\alpha_1^O-\alpha_1^C, \delta_1^O-\delta_1^C, \ldots, \delta_n^O-\delta_n^C)$ is the vector of the astrometric residuals of $n$ observations with the observed values $O$ and the computed values $C$. The weight matrix $W$ is a $(2n \times 2n)$ matrix describing the uncertainties of the various measurements. We use a weight matrix with the inverse standard errors $1/\sigma^2$ as diagonal elements. The standard deviations are given in Table\;\ref{Tab:StandardDev} \cite{farnocchia2015systematic}. An iterative solution with the least squares fit method can be achieved by adding 
\begin{equation}\label{Eq:LeastSquaresFit}
    \Delta\mathbf{A} = -(B^T W B)^{-1} B^T W \boldsymbol\nu
\end{equation}
to $\textbf{A}$, where $B=\partial \boldsymbol\nu / \partial \mathbf{A}$ is a $(2n \times 4)$ matrix of measurement equation coefficients. An appropriate initial attributable can be estimated by the first and last observations $\mathbf{A}_\mathrm{ini} = (\alpha_1,\delta_1,(\alpha_n-\alpha_1)/(t_n-t_1), (\delta_n-\delta_1)/(t_n-t_1))$. Note, that we include a light-time correction by subtracting $\varrho/c$ from the observed epochs for any propagation computation with $c$ as speed of light.

\begin{table}
\centering
\begin{tabular}{cc}
\hline             
Obs. code     & $\sigma_{\alpha,\delta}$ [arcsec] \\ \hline
568 & 0,15 \\
F51 & 0,20\\
H01 & 0,30 \\
E10, F65, J04, W84 & 0,40\\
291, 691, 950, G96 & 0,50 \\ 
W85, W86, W87 & 0,60 \\
H21 & 0,70 \\
G45, K91, K92, K93 & 0,80 \\
Q63, Q64, V37 & 0,80 \\
703 & 1.00 \\
\hline                
\end{tabular}
\caption {Astrometric standard deviation for a number of important observatories as used in \cite{farnocchia2015systematic}. We use $\sigma_{\alpha,\delta}=1\,\mathrm{arcsec}$ for non-listed observatories.} \label{Tab:StandardDev}
\end{table}

\begin{figure*}
    \centering
    \includegraphics[width=1.00\textwidth]{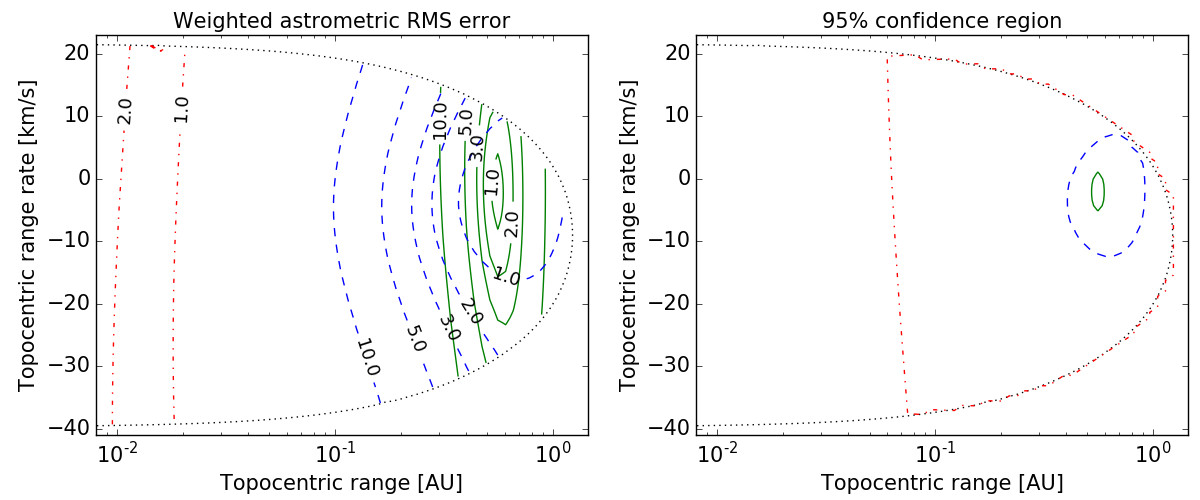}
    \caption{The left plot shows the weighted astrometric RMS error of the object P10LiAR. The red dash-dotted contour lines correspond to a tracklet with 3 observations by station F51, spanning a time period of $37\;\mathrm{min}$. A tracklet of $6$\;observations, 3\;additional observations done by station 568, covering $47\;\mathrm{h}$ is shown by blue dashed contour lines. The green solid contour lines are the result of a tracklet with 9 observations within $49\:\mathrm{h}$, where the latest 3 follow-ups were done by station 691. The right plot shows the corresponding 95\% confidence regions. The black dotted line is the boundary between elliptic and hyperbolic orbits.}
	\label{fig:systematicranging_P10LiAR}
\end{figure*}

\begin{figure*}[p!]
    \centering
    \includegraphics[width=1.00\textwidth]{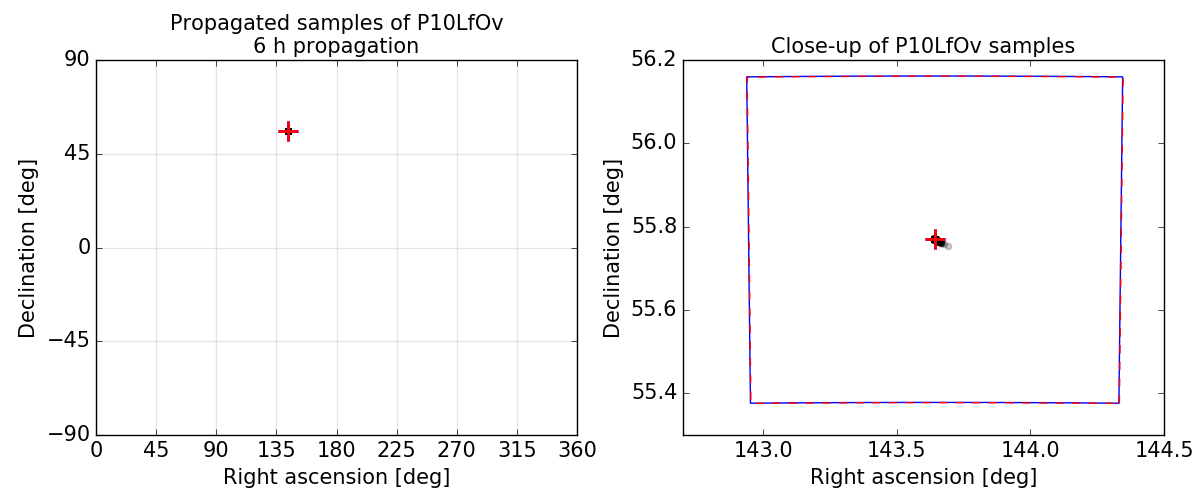}
    \includegraphics[width=1.00\textwidth]{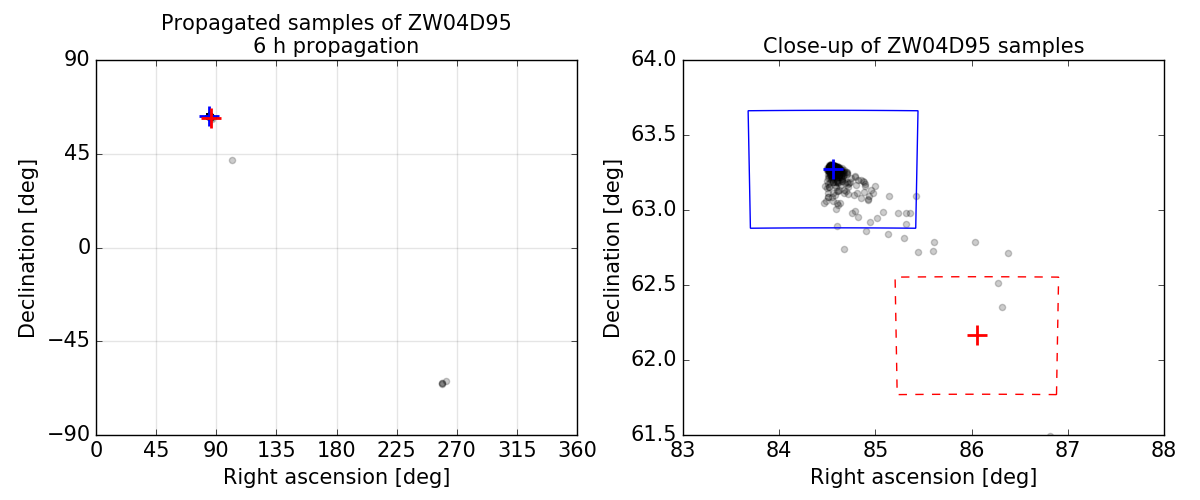}
    \includegraphics[width=1.00\textwidth]{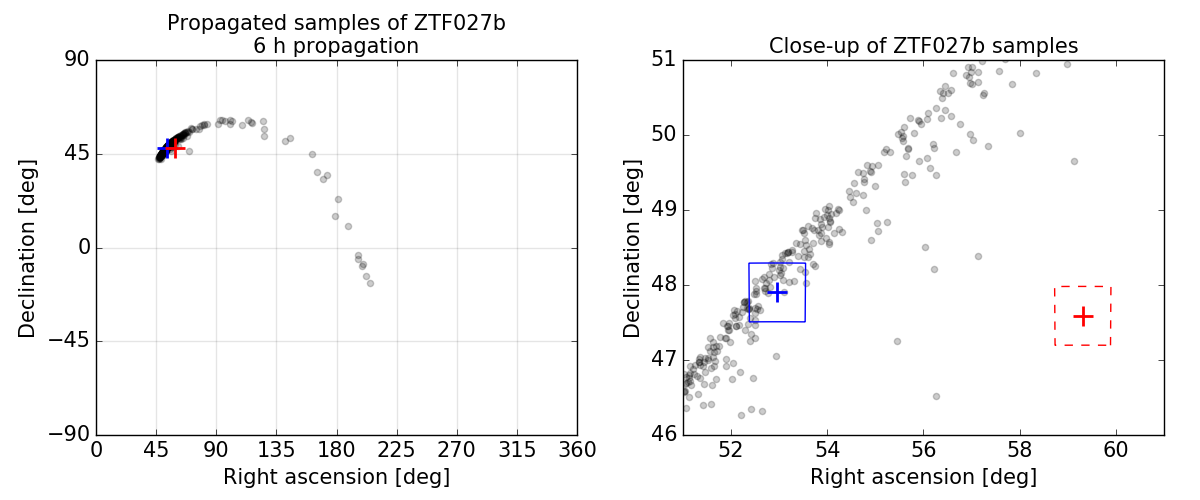}
    \caption{The figure shows the sample clouds (black dots) of P10LfOv (top), ZW04D95 (center) and ZTF027b (bottom) for a $6\;\mathrm{h}$ propagation after the last observation. The left plots are all-sky overviews, while the right plots are close-ups. The promising pointing computations for the mean (red) and the median (blue) are marked as crosses. It can be seen that it is important to use the median position, rather than the mean, to compute the promising pointing. An example field of view (FOV) of $47\;\mathrm{arcmin}$, corresponding to ESA's Optical Ground Station (J04), is shown.}
	\label{fig:followup_12h}
\end{figure*}

\begin{figure*}
    \centering
    \includegraphics[width=1.00\textwidth]{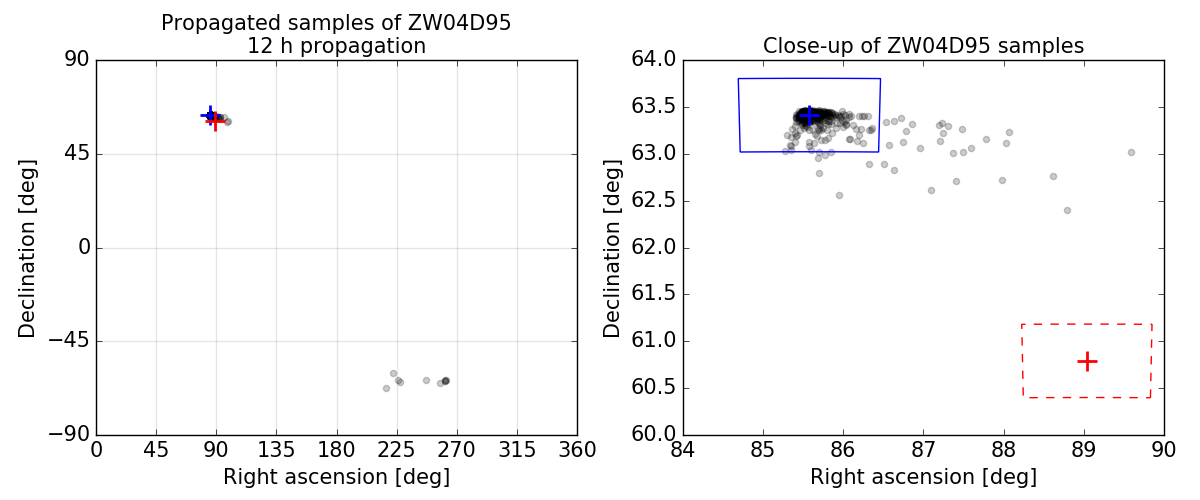}
    \caption{The figure shows the sample cloud of ZW04D95 (black dots) for a $12\;\mathrm{h}$ propagation after the last observation. The left plot is an all-sky overview, while the right plot is a close-up. The promising pointing computations for the mean (red) and the median (blue) are marked as crosses. An example field of view (FOV) of $47\;\mathrm{arcmin}$, corresponding to the Optical Ground Station (J04), is shown.}
	\label{fig:followup_24h}
\end{figure*}

For each grid point a weighted astrometric root mean square (RMS) error can be computed, indicating the quality of the orbit fit:
\begin{equation}\label{Eq:RMS}
    \mathrm{RMS} = \sqrt{\frac{1}{2n} \boldsymbol\nu^T W \boldsymbol\nu}
\end{equation}
An example can be seen in left plot of Figure\;\ref{fig:systematicranging_ZW04D95} for the object with the temporary designation ZW04D95\footnote{https://www.minorplanetcenter.net\label{FN:mpc}}. The black dotted line corresponds to the boundary between elliptic and hyperbolic orbits. The blue solid contour lines show the weighted astrometric RMS error of the best-fit solution. To validate our results we compare them with the contour map (red dashed contour lines) of JPL's Scout website\footnote{https://cneos.jpl.nasa.gov/scout\label{FN:scout}}, provided by D. Farnocchia (personal communication, 2018). Both RMS errors are computed by using the same astrometric standard deviations $\sigma_{\alpha,\delta}$. The two results match very well, except for approaching orbits close to Earth. In this region the gravity of Earth and Moon gets significant, yet for performance reasons we only use a Kepler propagator assuming unperturbed heliocentric orbits. Hence, this difference is expected. The perturbations are important for an accurate orbit determination, which is why we will include them in future work, though they can be neglected for the goal of this paper.

We compute a posterior probability density $f_\mathrm{post}$ for topocentric range and range rate, by multiplying the error function $f_\mathrm{err}$ of the normally distributed observation errors with a prior probability density function $f_\mathrm{prior}$, as given by \cite{farnocchia2015systematic}: 
\begin{align}
    & f_\mathrm{post} \propto f_\mathrm{err} f_\mathrm{prior} \label{Eq:PDF} \\
    & f_\mathrm{err}= \exp\left(-\frac{1}{2}\boldsymbol\nu^T W \boldsymbol\nu\right)\\
    & f_\mathrm{prior} = \varrho^2 f(H) = \varrho^{2-5\eta}
\end{align}

with a power law size distribution $f(H)\propto\varrho^{-5\eta}$ and $\eta$ ranging between $0.35$ and $0.47$ \cite{farnocchia2015systematic}. We select for our computations $\eta=0.41$. For the probability density computation only heliocentric bound orbits with eccentricity $e<1$ are considered, while the probabilities for unbound orbits are set to zero. The probability density is plotted on the right side of Figure\;\ref{fig:systematicranging_ZW04D95} as blue solid lines. By summing up the $f_{post}$ values, starting with the lowest probabilities, to 5\% of the total grid sum, we determine a 95\% confidence region of the ($\varrho,\dot{\varrho}$)-space. This region is enclosed by the magenta dashed line.

\section{Monte Carlo samples}

To predict the distribution, where the object may be found at an arbitrary epoch, we generate Monte Carlo samples, following \cite{farnocchia2015systematic}:

\begin{enumerate}
\item Randomly drawing a topocentric range $\varrho_k$ and range rate $\dot{\varrho}_k$ within the grid limits;
\item Randomly picking a posterior probability density $\gamma_k$ between zero and the upper limit of the maximum computed value of the grid;
\item Computing the attributable $\mathbf{A}_k(\varrho_k,\dot{\varrho}_k)$ and posterior probability density $f_{\mathrm{post},k}(\varrho_k,\dot{\varrho}_k)$;
\item Continuing if $f_{\mathrm{post},k} \ge \gamma_k$, otherwise restarting with step 1;
\item Adding Gaussian noise, defined by the covariance matrix $\Gamma_k = ({B_k}^T W B_k)^{-1}$, to $\mathbf{A}_k(\varrho_k,\dot{\varrho}_k)$;
\end{enumerate}

For a better resolution at small topocentric ranges, we sample the range in a logarithmic scale. In this case, the density function of Equation\;\ref{Eq:PDF} has to be multiplied with\;$\varrho$. We generate 500 Monte Carlo samples in Figure\;\ref{fig:systematicranging_ZW04D95}, where as expected most of them are located inside the magenta dashed 95\% confidence region. Starting from the generated topocentric ranges, range rates and computed attributables, the orbital elements can be derived and processed, e.g. to propagate the samples and check for impacts on Earth to estimate an impact probability.

\section{Change of RMS error and 95\% confidence region after follow-up observations}

We investigate the change of RMS error and 95\% confidence region after follow-up observations. To this aim, we repeat the computations of the object P10LiAR with different number of observations. The first tracklet consists of 3\;observations, the initial discovery, by the station Pan-STARRS\;1 (F51), spanning a time period of $37$\;min. The second tracklet includes 3\;follow-ups taken from Mauna Kea (568), leading to a total of 6\;observations covering $47$\;h. Finally we use 9\;observations within $49$\;h, including data from the Steward Observatory (691). The result can be seen in Figure\;\ref{fig:systematicranging_P10LiAR}, where the red dash-dotted lines corresponds to 3 observations, the blue dashed lines to 6 observations and the green solid lines to 9 observations. The left plot shows the weighted astrometric RMS error and the right plot shows the 95\% confidence region. One can see that the region with a good fit of the observations shrinks with more data taken over a larger time span and from independent data sources. As a consequence, the 95\% confidence region decreases too.

\section{Promising follow-up pointing} \label{Sec:bestfollowuppointing}

We want to find a promising right ascension $\alpha_\mathrm{f}$ and declination $\delta_\mathrm{f}$ to point at for follow-up observations. To this aim, we propagate the samples to a certain epoch. 

\begin{figure*}
    \centering
    \includegraphics[width=1\textwidth]{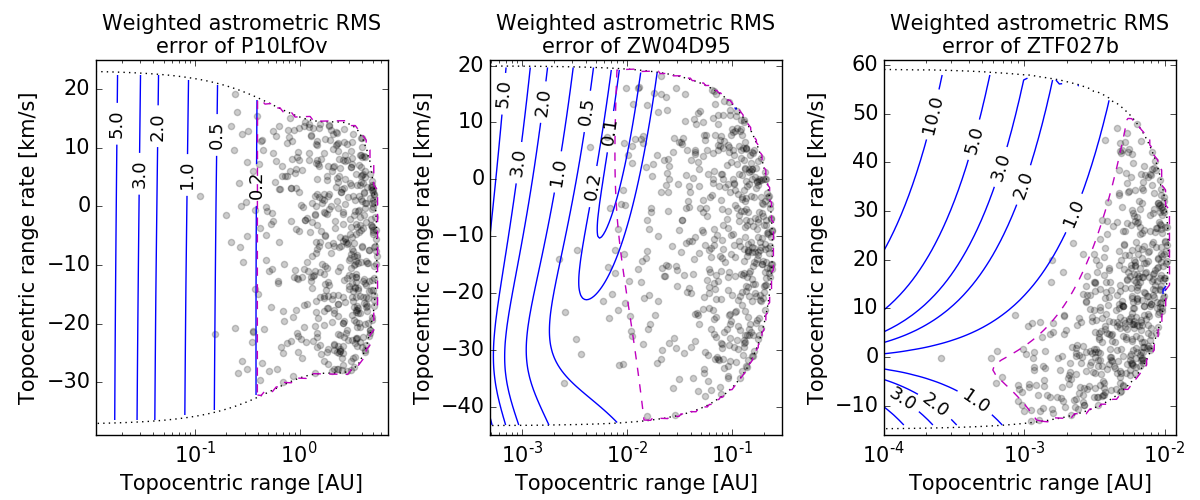}
    \caption{The figure shows the weighted astrometric RMS error (blue solid contour lines) and the 95\% confidence region (magenta dashed lines) of the object P10LfOv (left), ZW04D95 (center) and ZTF027b (right). 500 Monte Carlo samples (black dots) are generated for each object. The black dotted lines correspond to the boundary between elliptic and hyperbolic orbits.}
	\label{fig:systematicranging_show}
\end{figure*}

A very simple approach is to compute the arithmetic means of $\alpha$ and $\delta$ of the propagated samples. This can be seen in Figure\;\ref{fig:followup_12h} as a red cross, where we propagated 500 random samples (black dots) of P10LfOv, ZW04D95 and ZTF027b to $6$\;h after their last observation. To keep the computation general, we use the geocenter as reference for the computation instead of a specific station. An example field of view (FOV) of $47$\;arcmin, corresponding to ESA's Optical Ground Station (J04), is shown as dashed frame around the cross. While for P10LfOv the sample cloud is centered very well, the mean of ZW04D95 is already shifted to a position where only 2 out of 500 samples are located in the FOV. This shift is the result of a few large outliers, which occur for samples that already came close to Earth and hence moved to the other side of the sky. The shape of the ZTF027b samples is not a roundish cloud anymore but very stretched and quite a few outliers can be seen. This shape arises from the computed posterior probability distribution and the resulting samples being placed in orbits closer to Earth than the samples of the other objects, which is shown in Figure\;\ref{fig:systematicranging_show}. As a result of the different ranges, the apparent velocities of the samples differ too, where P10LfOv has the lowest apparent velocities and ZTF027b the highest. Those high apparent velocities lead to the spread of the cloud. For ZTF027b, the mean computation would suggest a bad pointing far off the sample line. If the samples are propagated $12$\;h instead of $6$\;h, as shown in Figure\;\ref{fig:followup_24h} for ZW04D95, the cloud spreads too, leading to an even more unfavorable pointing recommendation.

Another approach is to compute the median of all $\alpha$ and $\delta$, which can be seen as a blue cross with solid frame in Figure\;\ref{fig:followup_12h} and Figure\;\ref{fig:followup_24h}. Similar to the mean, the median centers the small cloud of P10LfOv very well, why one cannot see a notable difference of the FOVs. In contrast, the median focuses very well on the $6$\;h and $12$\;h propagated samples of ZW04D95. For ZTF027b, the median places the pointing recommendation on the sample line, yet it centers not necessarily on the densest region. However, since the samples are widely spread and the sample density inside the FOV is low anyways, not finding a promising pointing for objects like ZTF027b will not make a big difference for the later computation of the detection probability. To rediscover ZTF027b after $6$\;h, an observation campaign or a piece of good luck is needed.

It can be seen that using the mean results in physically unrealistic solutions, the median gives better results. This is due to the fact that errors are not Gaussian-distributed, similar to what was shown in \cite{schmidt2019analysis}. Therefore, for the rest of the paper, we will use the median to compute the promising pointing.

\section{Follow-up pointing for chosen telescope}
\label{Sec:PointingTelescope}

If a specific telescope is chosen for the follow-up observations, not all samples might be detectable as,

\begin{enumerate}
\item the topocentric range $\varrho$ differs for each sample and the resulting apparent magnitude $m$ might be too faint for a specific telescope with a given limiting visual magnitude.
\item depending on the telescope latitude, longitude and follow-up observation epoch, some of the samples could have $\alpha$ and $\delta$, which are below the horizon.
\end{enumerate}

Those undetectable samples are not considered in determining the median values, but only the detectable ones.

\begin{figure*}[p]
    \centering
    \includegraphics[width=1.00\textwidth]{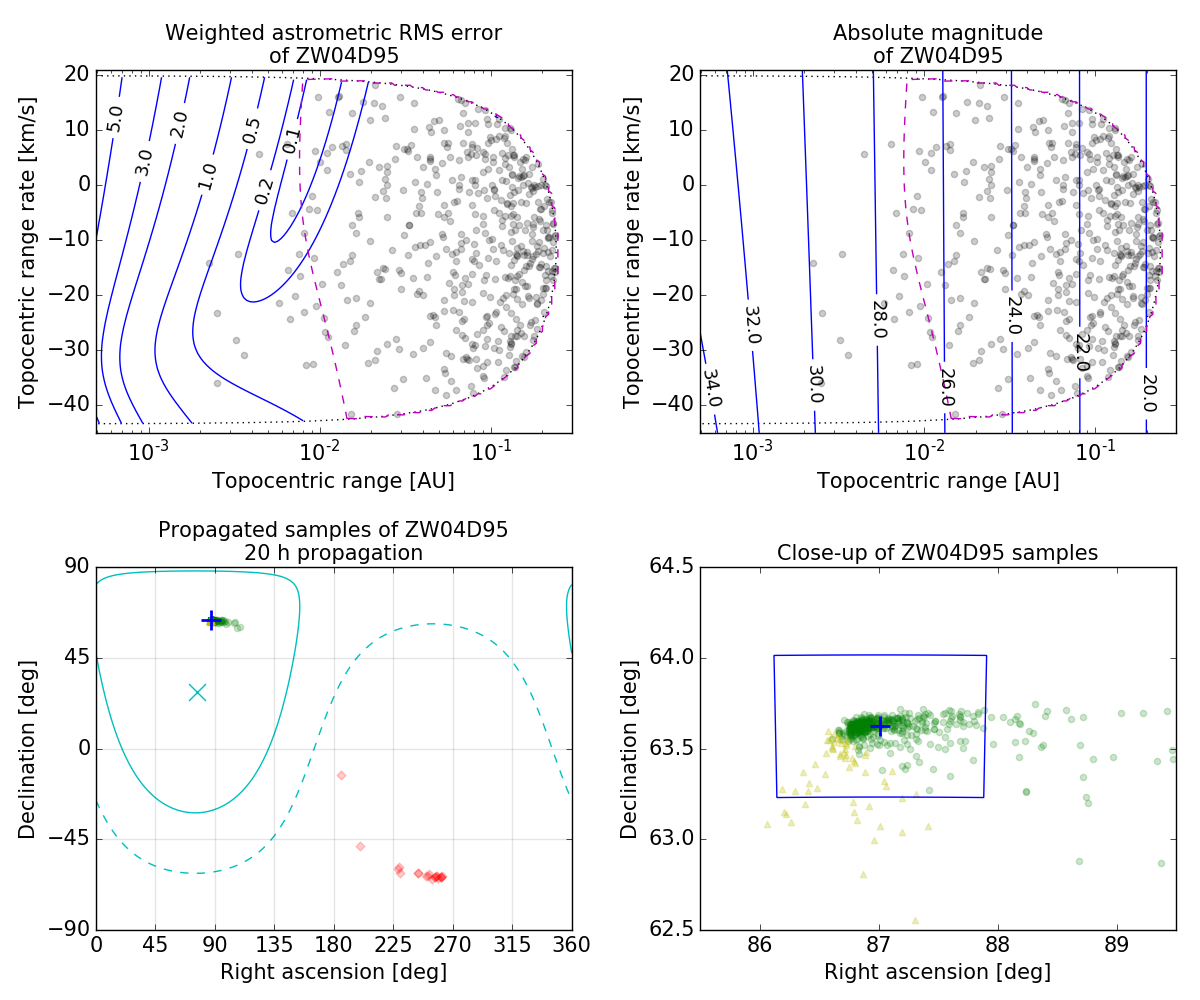}
    \caption{The upper plots show the weighted astrometric RMS error and the absolute magnitude $H$ of the topocentric range - range rate grid for the object ZW04D95. In both plots the 95\% confidence region (magenta dashed) and 500 samples are placed. The black dotted line is the boundary between elliptic and hyperbolic orbits. The lower plots show the samples for a $20\;\mathrm{h}$ propagation, as seen from ESA's Optical Ground Station (J04). The cyan X corresponds to the zenith of the observatory, the cyan dashed line to the horizon and the cyan solid line to $30\;\mathrm{deg}$ elevation. We distinguish between samples below $30\;\mathrm{deg}$ elevation (red diamonds), to faint samples (yellow triangles) and visible samples (green dots). The computed pointing, is shown with a blue cross, surrounded by the FOV projection of $47\;\mathrm{arcmin}$. 356 visible samples out of 500 are located inside the FOV, leading to a detection probability $p \approx 0.71$.}
	\label{fig:propability_20h_ZW04D95}
\end{figure*}

\begin{figure*}
    \centering
    \includegraphics[width=0.64\textwidth]{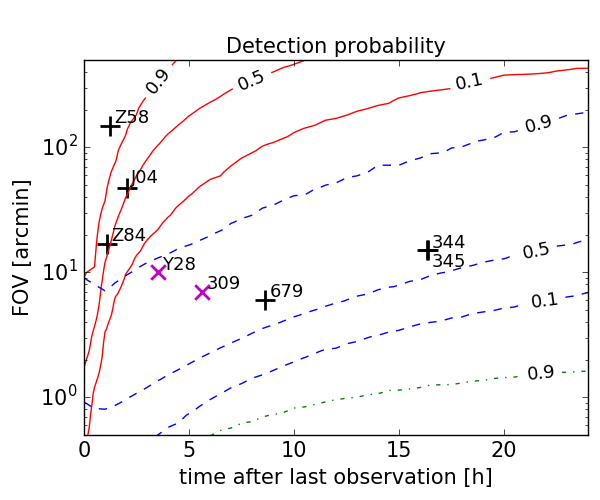}
    \caption{The plot shows the contour lines of the detection probability for ZTF027b (red solid) with a high apparent velocity, ZW04D95 (blue dashed) with an intermediate apparent velocity and P10LfOv (green dash-dotted) with a low apparent velocity, depending on the time after the last observation and the field of view (FOV). The propagation was done for a theoretical station at the geocenter. The crosses show the stations of Table\;\ref{Tab:fieldofview}, placed with their specific FOV and the longitudinal difference in hours between the station and the future Flyeye telescope at Monte Mufara. Black crosses correspond to observatories of the northern hemisphere, magenta X to the southern hemisphere.}
	\label{fig:probability_timeFOV}
\end{figure*}

The absolute magnitude $H$ in the H-G magnitude system \cite{bowell1989application} can be computed from the apparent magnitude $m$ by
\begin{align}
\label{Eq:MagnitudeConversion}
    & H = H(\varphi) + 2.5 \log\big( (1-G)\Phi_1(\varphi) + G\Phi_2(\varphi) \big) \\
    & H(\varphi) = m - 5 \log_{10}(d \varrho) \\
    & \Phi_i = W \Phi_{i\mathrm{S}} + (1-W) \Phi_{i\mathrm{L}}\\
    & W = \exp\left(-90.56 \tan(\varphi/2)^2\right)\\
    & \Phi_{i\mathrm{S}} = 1- \frac{C_i \sin(\varphi)}{0.119+1.341 \sin(\varphi) - 0.754 \sin(\varphi)^2}\\
    & \Phi_{i\mathrm{L}} = \exp\left(-A_i \tan(\varphi/2)^{B_i}\right)
\end{align}
and vice versa, where $i=1,2$. The function $H(\varphi)$ is called reduced magnitude with the phase functions $\Phi_i$. $\varphi$\;is the phase angle and $d$ corresponds to the heliocentric distance of the sample. The corresponding constants are $A_1 = 3.332$, $B_1 = 0.631$, $C_1 = 0.986$, $A_2 = 1.862$, $B_2 = 1.218$ and $C_2 = 0.238$. As we do not have observed phase curves for newly discovered NEOs, we assume a slope parameter $G=0.15$ for all objects.

Similar to the attributable fit of Equation\;\ref{Eq:CostFunction}, we compute the best fit $H_k$ from the photometric data by a least squares fit for each sample, where we consider the previously found best fit attributable. Finally the $m_k$ of the samples depend on the propagation epoch.

An example of the $H$ computation can be seen in the upper right plot of Figure\;\ref{fig:propability_20h_ZW04D95}, where we computed the absolute magnitude of ZW04D95 as function of $\varrho$ and $\dot{\varrho}$.

\section{Detection probability for a chosen telescope}

To compute the number of detected samples by a follow-up observation, we map the samples with their right ascension $\alpha$ and declination $\delta$ on the field of view of a specific telescope by \cite{montenbruck2000astronomy}:
\begin{eqnarray}\label{Eq:ProjectionX}
    &X = - \frac{\cos(\delta) \sin(\alpha-\alpha_\mathrm{f})} {\cos(\delta_\mathrm{f})\cos(\delta)\cos(\alpha-\alpha_\mathrm{f}) + \sin(\delta_\mathrm{f})\sin(\delta)}\\
    &Y = - \frac{\sin(\delta_\mathrm{f})\cos(\delta)\cos(\alpha-\alpha_\mathrm{f}) - \cos(\delta_\mathrm{f})\sin(\delta)} {\cos(\delta_\mathrm{f})\cos(\delta)\cos(\alpha-\alpha_\mathrm{f}) + \sin(\delta_\mathrm{f})\sin(\delta)}
\end{eqnarray}
where $\alpha_\mathrm{f}$ corresponds to the right ascension center of the follow-up observation and $\delta_\mathrm{f}$ the declination center. If a sample lies within the boundaries of the FOV by checking {$\vert X \vert \le \arctan(\mathrm{FOV}/2)$} and {$\vert Y \vert \le \arctan(\mathrm{FOV}/2)$}, the sample is found. Depending on the specific telescope and CCD properties, further mapping, e.g. due to a rotated CCD, must be implemented before the check.

Only the visible samples, which are used to determine the center of observation, have to be accounted for. Dividing the sum of found samples by the full sample number, including the too faint and below horizon objects, leads to an overall detection probability $p$ of a single observation by a specific telescope at a certain epoch. To get a reliable result, a large enough Monte Carlo sample sequence has to be used.

\begin{table*}[t]
\centering
\begin{tabular}{ccc}
\hline             
Obs. code     & Telescope name & FOV [arcmin] \\ \hline
J04 & 1.0 m Optical Ground Station at Teide & 47 \\
Z84 & 0.8 m Schmidt telescope & 17 \\
Z58 & 0.56 m Test-Bed Telescope & 149 \\
309 & 8.2 m Very Large Telescope & 7 \\
344 & 1.8 m BOAO telescope & 15 \\
345 & 0.6 m SOAO telescope & 15 \\
Y28 & 1.0 m OASI telescope & 10 \\
679 & 2.12 m telescope at San Pedro Martir & 6 \\\hline                
\end{tabular}
\caption {Set of follow-up telescopes with their field of view (FOV).} \label{Tab:fieldofview}
\end{table*}

An example can be seen in Figure\;\ref{fig:propability_20h_ZW04D95}, where the promising pointing for the object ZW04D95 is shown in the lower plots. We chose ESA's Optical Ground Station (J04) with a FOV of $47$\;arcmin as follow-up observatory and $20$\;h after the last observation as observation epoch. Instead of simply considering the horizon as observation limit, we set the limit to $30$\;deg elevation (cyan solid line), where one would expect good observation data. The zenith of the station is indicated by the cyan X and the horizon by the cyan dashed line in the lower left plot. For demonstration reasons, the limiting magnitude is set to $m=18.5$. The blue cross marks the computed promising pointing, surrounded by the projected FOV. Green dots show detectable objects while red diamonds are below the horizon and yellow triangles are too faint. Out of 500 samples, 356 observable ones match with the FOV projection, leading to a detection probability of $p(20\;\mathrm{h},\mathrm{J04})\approx0.71$.

\section{Detection probability evolution}

As we have already noticed in Chapter\;\ref{Sec:bestfollowuppointing}, the sample cloud spreads with time. To investigate this effect on the detection probability, we neglect the other influences on $p$, i.e. the horizon and the limiting magnitude, which both depend on the observatory. The propagated right ascensions and declinations are once more computed from the geocenter, to obtain a general, non-station dependent solution. 

The detection probability of a certain object depends now on the time after the last observation and the field of view, which is shown in Figure\;\ref{fig:probability_timeFOV} for the same 3\; examples as in Chapter\;\ref{Sec:bestfollowuppointing}. The main difference between the observed asteroids is their apparent velocity: P10LfOv (green dash-dotted) is slow, ZW04D95 (blue dashed) is intermediate and ZTF027b (red solid) is fast. As expected, $p$ shrinks with increasing time, as the sample cloud spreads, and with decreasing FOV. As observed before in Figure\;\ref{fig:followup_12h}, we know that the sample cloud of P10LfOv is very narrow after $6$\;h, which is why even after $12$\;h the spread has not a large effect on the detection probability for FOVs of the order of $1$\;arcmin. A clear difference can be seen for ZW04D95, where $p$ ranges from less than 0.1 to more than 0.9 in the given boundaries. The most challenging example is ZTF027b, where the detection probability decreases very quickly within 10\;h even for large FOVs of several hundreds of arcmin to $p<0.5$. For FOVs of the order of $10$\;arcmin, a probability of less than 10\% is already reached within a few hours.

As we expect the Flyeye telescope to find many new NEOs, we investigate the chances of successful follow-ups of those objects. To this aim, a set of data points representing several follow-up telescopes, given in Table\;\ref{Tab:fieldofview}, is placed in Figure\;\ref{fig:probability_timeFOV}. We distinguish between stations of the northern hemisphere (black cross) and southern hemisphere (magenta X). The abscissa represents the longitude difference between the future Flyeye telescope longitude at Monte Mufara and the follow-up telescope longitude in hours. Therefore if an object similar to the given examples is detected by the Flyeye telescope, we can estimate the detection probability by certain stations at their location. 

Among our chosen set of stations, the likelihood of observing P10LfOv and ZW04D95-like objects is very high. In contrast, ZTF027b-like objects are a bigger challenge as there is only a single considered station with $p>0.9$, the Test-Bed Telescope (Z58). Two more observatories of the northern hemisphere have chances close to 50\%. The other stations are clearly below 10\%. Based on Figure\;\ref{fig:probability_timeFOV} we can estimate the chance of observing an object and decide for a station with high enough detection probability to do follow-ups. For objects like ZTF027b, Z58 will be one of the important follow-up observatories of the Flyeye telescope. However, the epoch of the real follow-up observation is not strictly limited to the longitude difference of the locations, but might be much earlier. This would lead to better $p$ for the stations.

Please note that the object might be below the horizon due to not ideal station latitude, which is only marginally considered by distinguishing between northern and southern hemisphere stations. As the Flyeye telescope will be located at Monte Mufara, Italy, telescopes of the northern hemisphere will have to do the early follow-ups to keep track of the new NEOs.

\section{Conclusions}

We created a software, based on the systematic ranging technique, scanning the space of topocentric range and range rate of a tracklet for the best-fitting orbits. Using a prior probability density function, we get a probability distribution, which is used to create a set of Monte Carlo generated samples. Those samples can be propagated to an arbitrary epoch, leading to a right ascension and declination distribution, which can be used by observers of any station to select a preferred telescope pointing to find the object. Finally, using the field of view of a chosen telescope, the detection probability for a pointing to the median of the Monte Carlo samples. With this information, observers quickly get an impression of the success chance of the scheduled follow-up.

We also investigated the evolution of the weighted astrometric RMS error and 95\% confidence regions for a typical NEO detection. Both shrink with additional data, showing the importance of follow-ups to have a longer arc for orbit determination. From Figure\;\ref{fig:followup_12h} and Figure\;\ref{fig:followup_24h}, one can see a large difference of the sample clouds. In particular NEOs close to Earth have elongated, spread clouds due to the high apparent velocities. Those objects are a challenge to follow-up, as their detection probability decreases quickly with time, shown in Figure\;\ref{fig:probability_timeFOV}. Expecting similar tracklets from the Flyeye telescope in the future leads to the conclusion that a number of stations, most notably the Test-Bed Telescope (Z58) in Cebreros, Spain, will have the important task to do the follow-ups of NEOs close to Earth, as they might be lost otherwise. In addition, such close objects could be impactors, emphasizing the urgency of observations to estimate a more precise impact probability and predict the potential impact locations. To this aim, the software must be fully automatized and will have to run immediately after the discovery of a new object, notifying observers of the discovery of a critical NEO.

\section*{Acknowledgments}
We thank Davide Farnocchia for assistance with the systematic ranging technique and providing his data for validation.

This research has made use of data and/or services provided by the International Astronomical Union's Minor Planet Center.


\end{document}